# Fast Stable STAP Algorithms Based on Feedback Orthogonalization

Vasily A. Khlebnikov[1] and Kristian Zarb Adami[2]

*Department of Astrophysics, University of Oxford*
*Denys Wilkinson Building, Keble Road, Oxford, OX1 3RH, United Kingdom*
[1] w.khlebnikov@ieee.org
[2] kza@astro.ox.ac.uk

*Abstract*— **The aim of this paper is to present a new fast-convergent numerically stable space-time adaptive processing (STAP) algorithm derived using a novel technique of feedback orthogonalization. The main advantages of this approach lie in its perfected stability to computational errors and faults which makes its real-time implementation on substantially faster and cheaper regular fixed-point processors possible.**

## I. Introduction

Most STAP algorithms designed for effective on-line application in fields such as radar, sonar, communications, seismology, radioastronomy, and medical imaging are required to have an extremely rapid rate of actual convergence. One way to speed up the real-time adaptive processing is to take advantage of faster cost-effective fixed-point parallel computing systems.

Due to the fact that the optimum weights associated with such criteria as the LMS, maximum SINR, and MVDR differ from the optimum Wiener solution only by a scale gain factor and have equal output SINR's, we consider an adaptive space-time filter optimized to maximize the SINR at its output [1].

In terms of computational structure, the quadratic optimisation STAP algorithms can be broadly classified under asymptotically convergent iterative and finite convergent direct categories. Since the iterative LMS-type algorithms implementing local searching optimization methods are unacceptably slow in most practical scenarios [2], we focus our attention on the finite convergent algorithms based on direct methods of linear algebra.

With respect to the number of iterations a numerical method requires to achieve an acceptable approximation to the optimum solution, the direct algorithms based on various types of matrix decomposition and recurrent inversion are considered to be very fast because under conditions of infinite precision arithmetic they converge in *2K* iterations, where *K* is the filter dimension [3]. However, numerous results of convergence analyses [2], [4-7] have shown that the direct algorithms are explosively divergent in regular finite precision environments necessitating their implementation on slower expensive floating-point processors.

The underlying cause of numerical instability of most direct STAP algorithms is in the strictly sequential feedback-less processing structure that results in uncontrollable propagation and accumulation of numerical errors in finite precision computing systems.

The main goal of this paper is to present a new fast-convergent algorithm for real-time STAP designed to be perfectly robust in terms of processing errors and faults that provides its real-time implementation on faster cost-effective regular fixed-point microprocessors. The new approach to the design is based on a technique of numerical errors propagation suppression applied to a direct computation method and realised by means of controlled massive parallel error-correcting feedback loops.

With regards to stochastic quadratic optimisation, this work can be considered as an approach to filling a gap between slow-convergent but stable gradient-based searching algorithms featured with parallel error-correcting feedback loops and fast-convergent but numerically instable direct methods, which are known to be without effective mechanisms of feedback error-correction.

The rest of the paper is organized as follows. Section II presents the conjugate direction decomposition (CDD) of the optimum filter weights. In Section III, we derive a new stochastic fast-convergent numerically stable STAP algorithm that computes the metric-orthogonal conjugate-direction basis using a new version of the modified Gram-Schmidt orthogonalization (MGSO) method. Results of numerical simulations for several sizes of fixed-point binary data/operation representation are demonstrated in Section IV. Brief conclusions are summarized in Section V.

## II. Background

In order to maximize the SINR at the space-time filter output for a given amount of input samples, we have to solve the generalized least-squares (GLS) problem [8]

$$\min_{\mathbf{w}_n} \| \mathbf{v}_S - \mathbf{R}_{Xn}\mathbf{w}_n \|^2, \quad n = 1,2,... \tag{1}$$

where $\boldsymbol{w}_n$ is the *K*-vector of filter weights, $\boldsymbol{v}_S$ is the reference *K*-vector and $\mathbf{R}_{Xn}$ is an estimator of the interference-plus-noise covariance matrix obtained using *n* input sampled *K*-vectors $\mathbf{x}_i$.

Using the singular value decomposition (SVD), the GLS solution can be represented as [8]

$$\mathbf{w}_n^+ = \mathbf{R}_{Xn}^+\mathbf{v}_S = \mathbf{U}_n\mathbf{\Lambda}_n^+\mathbf{U}_n^H\mathbf{v}_S = \sum_{i=1}^{r_n}\frac{\mathbf{u}_{ni}\mathbf{u}_{ni}^H}{\lambda_{ni}}\mathbf{v}_S = \sum_{i=1}^{r_n}\alpha_{ni}\mathbf{u}_{ni} \tag{2}$$

where $R_{Xn}^+$ is the Moore-Penrose pseudoinverse $K\times K$ matrix, $\mathbf{U}_n$ is the $K\times r_n$ matrix of singular vectors and $\mathbf{\Lambda}_n$ is the real diagonal matrix of the corresponding singular values arranged in descending order. The number of non-zero singular values, $r_n$ – the effective rank of matrix $\mathbf{R}_{Xn}$ – is always bounded, i.e.

$$r_n \le \min(n, K) \tag{3}$$

According to the solution (2), the space-time adaptive filter is a two-stage $K$-input signal processor that contains a *decorrelation filter* associated with matrix $\mathbf{U}_n$ followed by a $K$-input *optimum linear combiner*.

The STAP algorithms developed using the EVD/SVD approach are known to be unsuitable for implementation on real-time fixed-point processors due to the requirements of arithmetic precision and numerical complexity [9-11].

In this paper, we suggest a new factorisation of the optimum GLS solution using an alternative vector basis that allows us to derive a new family of fast convergent adaptive algorithms with improved numerical stability, better fault tolerance, and lowered computational complexity.

Linear algebra states [12] that the optimum weights in (2) can also be represented by making use of a *basis of conjugate directions* as

$$\mathbf{w}_n^+ = \mathbf{S}_n \Delta_n^+ \mathbf{S}_n^H \mathbf{v}_S = \sum_{i=1}^{r_n} \frac{\mathbf{s}_{ni}\mathbf{s}_{ni}^H}{\delta_{ni}} \mathbf{v}_S = \sum_{i=1}^{r_n} \beta_{ni}\mathbf{s}_{ni} \tag{4}$$

where $\mathbf{S}_n$ is a $K{\times}r_n$ column matrix containing $r_n$ conjugate directions which are orthogonal only with respect to the matrix $\mathbf{R}_{Xn}$, i.e.

$$\mathbf{s}_{ni}\mathbf{R}_{Xn}\mathbf{s}_{nj}^H = 0, \quad i \ne j, \quad i, j \in [1, r_n] \tag{5}$$

and

$$\Delta_n = \operatorname{diag}\{\delta_{ni}\}_1^{r_n} = \operatorname{diag}\{\mathbf{s}_{ni}^H \mathbf{R}_{Xn}\mathbf{s}_{ni}\}_1^{r_n} \tag{6}$$

contains the $\mathbf{R}_{Xn}$-norms of conjugate directions arranged in diminishing order.

From (5) it is evident that the matrix $\mathbf{S}_n$ decorrelates the filter input signals as well as the matrix $\mathbf{U}_n$ thus making the elements of the output vector $\mathbf{y}_n = \mathbf{S}_n^H \mathbf{x}_n$ mutually orthogonal, i.e. $\mathbf{R}_{Yn} = E\{\mathbf{y}_n \mathbf{y}_n^H\} = \mathbf{\Delta}_n$.

Although the conjugate directions are orthogonal only in the metric described by the input covariance matrix, the CDD offers one clear advantage over the EVD/SVD. Given a $K{\times}K$ nonnegative-definite matrix $\mathbf{R}_{Xn}$ and an arbitrary set of $K$ linearly-independent $K$-vectors, the conjugate basis can be computed in $K$ recursions utilizing simple stable inexpensive methods of vector orthogonalization.

## III. Adaptive Algorithm Based on the Modified Gram-Schmidt Orthogonalization

In the first part of this section, we introduce a new version of the modified Gram-Schmidt orthogonalization method (MGSO) to compute a metric-orthogonal vector basis with suppressed levels of errors accumulation and propagation. A novel fast stable STAP algorithm developed basing on the principle massive staggered feedback orthogonalization is presented in the second part.

### *A. Numerically Stable Metric-Orthogonalization Method*

Consider a $K{\times}M$ column matrix $\mathbf{S}_0$ containing an arbitrary set of $M$ linear-independent vectors ($M \le K$) in $C^K$ and $\mathbf{R}$ is a $K{\times}K$ full-rank matrix. Assume also that the vectors $\mathbf{s}_k$ are arranged in order of $\mathbf{R}$-norm decrease.

In order to construct such a set of $\mathbf{R}$-orthogonal vectors, we use the following recursive orthogonalization form that differs from the MGSO only within the metric of orthogonality

$$\mathbf{S}(i) = \mathbf{S}(i-1)\mathbf{H}(i), \quad i = 1, \dots, M-1 \tag{7}$$

where $\mathbf{S}(i)$ is the $K{\times}M$ matrix containing $i$+1 conjugate directions and $\mathbf{H}(i)$ is the $M{\times}M$ basis transformation matrix, which differs from the $M{\times}M$ identity matrix only in the $i$-th row whose non-zero elements are calculated as

$$h_{ik} = -\frac{\mathbf{s}_i^H \mathbf{R}\mathbf{s}_k}{\mathbf{s}_i^H \mathbf{R}\mathbf{s}_i}, \quad k = i+1, \dots, M \tag{8}$$

Similarly to the stabilized MGSO [13] the remaining vectors $\mathbf{s}_{i+1}, \dots, \mathbf{s}_M$ must be rearranged in order of $\mathbf{R}$-norm decrease to suppress the volumes of transferred numerical errors implying that $\|h_{ik}(i)\|^2 \le 1$ for all $i$ and $k$.

Thus the resulting set of $\mathbf{R}$-orthogonal vectors is

$$\mathbf{S} = \mathbf{S}_0 \prod_{i=1}^{M-1} \mathbf{H}(i) = \mathbf{S}_0 \mathbf{H} \tag{9}$$

where $\mathbf{H}$ is the $M{\times}M$ upper triangle matrix whose off-diagonal non-zero elements are limited in absolute value to unity.

From (9) it follows that if the initial vector system $\mathbf{S}_0$ is set to be the identity matrix columns in the case of $M = K$ then the final conjugate basis becomes $\mathbf{S} = \mathbf{H}$ having only $K(K+1)/2$ non-zero components.

### *B. CDD Based Adaptive Algorithm*

In most practical cases, the input covariance matrix $\mathbf{R}_{Xn}$ is estimated using the following well-known recurrent form

$$\mathbf{R}_{Xn} = (1 - 1/n)\mathbf{R}_{X(n-1)} + 1/n\,\mathbf{x}_n \mathbf{x}_n^H, \quad \mathbf{R}_{X0} = \mathbf{0}_K \tag{10}$$

According to the recurrent update in (10), the conjugate-direction basis for the decomposition in (3) is built using the stable version of metric-orthogonalization given above as

$$\mathbf{S}_n = (1 - 1/n)\mathbf{S}_{n-1} + 1/n\,\mathbf{S}_n \mathbf{H}_n, \quad \mathbf{S}_0 = \mathbf{I}_K \tag{11}$$

The structure of elementary $\mathbf{R}_{Xn}$-orthogonalization matrices $\mathbf{H}_n$ in (11) depends on the sample number $n$ and estimated effective rank $r_n$. For example, matrix $\mathbf{H}_1$ may have only one row containing $K$-1 non-zero elements in addition to its unit diagonal while matrix $\mathbf{H}_K$ may contain $r_n$ 'non-zero' rows.

The elements of an $i$-th non-zero row in matrix $\mathbf{H}_n$ are

$$\begin{aligned} h_{nik} &= -\frac{c_{nik}}{\sigma_{ni}^2}, \quad \text{if } \sigma_{ni}^2 \ge \sigma_{nk}^2 \\ h_{nik} &= 0, \quad \text{if } \sigma_{ni}^2 < \sigma_{nk}^2, \quad k = 1, \dots, r_n \end{aligned} \tag{12}$$

where

$$
\begin{aligned}
c_{nik} &= (1-\tfrac{1}{n_I})c_{(n-1)ik} + \tfrac{1}{n_I}\, y_{ni} y_{nk}^{*} \\
\sigma_{ni}^{2} &= (1-\tfrac{1}{n_I})\sigma_{(n-1)i}^{2} + \tfrac{1}{n_I}\, |y_{ni}|^{2} \\
y_{ni} &= \mathbf{s}_{ni}^{H}\mathbf{x}_{n}, \quad i,k = 1,\ldots,r_n
\end{aligned}
\tag{13}
$$

In (13), $n_I$ =2,3, … is the amount of sequential input samples that may be required to smoothen the estimates of the output second-order statistics at each recursion, $y_{ni}$ is the $i$-th output of the matrix $\mathbf{S}_n$ defined decorrelation filter, $c_{nik}$ is the averaged covariance between the $i$-th and $k$-th filter output and $\sigma_{ni}^2$ is the $i$-th filter output averaged power.

Finally, the adaptive MSINR weights in (4) become

$$
\mathbf{w}_n^{+} = \mathbf{S}_n \Delta_n^{+} \mathbf{S}_n^{H} \mathbf{v}_S = \sum_{i=1}^{r_n} \frac{\mathbf{s}_{ni}\mathbf{s}_{ni}^{H}}{\sigma_{ni}^{2}} \mathbf{v}_S = \sum_{i=1}^{r_n} \beta_{ni}\mathbf{s}_{ni} \tag{14}
$$

Equations (11) to (14) highlight how the use of the CDD has allowed us to replace direct metric-orthogonalization of input vectors $\mathbf{x}_n$ with regular orthogonalization applied to the decorrelation filter outputs $\mathbf{y}_n$. In thus way there is effective error control on the performance of adaptive space-time processing.

A computation schematic that summarises the design of the new fast stable STAP algorithm is presented in Table I.

TABLE I
CDD BASED FAST STABLE STAP ALGORITHM

♦ INITIALISATION

- transformation matrix: $\mathbf{H}_0 = \mathbf{I}_K$
- matrix of conjugate directions: $\mathbf{S}_0 = \mathbf{I}_K$

♦ ADAPTATION

**for** $n$ = 1,2, … **do**

- current amount of the CD's: $N_S = n$; **if** $n \geq K$ **then** $N_S = K$
- initialized covariation matrix at the decorrelator's outputs: $\mathbf{R}_{Y0} = \mathbf{0}_K$
- smoothing the decorrelator's output covariance matrix:

**for** i = 1, … , $n_I$ **do**

$$\mathbf{R}_{Yn} = (1-\tfrac{1}{i})\mathbf{R}_{Y(n-1)} + \tfrac{1}{i}\,\mathbf{S}_{n-1}^{H}\mathbf{x}_{n_I(n-1)+i}\mathbf{x}_{n_I(n-1)+i}^{H}\mathbf{S}_{n-1}$$

**end** $i$

- finding the current index to the next maximum output power:

$i_{Sn}$ = argmax$\{r_{Ykk}\}$, k = 1, … , K, k ≠ $i_{Sn\text{-}1}$, $i_{Sn\text{-}2}$, …

- updating matrix $\mathbf{H}_n$:

**for** $m = i_{S1}$, … , $i_{Sn}$ **do**

**for** $k$ = 1, … , $K$ **do**

**if** $r_{Ymm} \geq r_{Ykk}$ **then** $h_{nmk} = -\, r_{Ynmk}/r_{Ynmm}$ **else** $h_{nmk} = 0$;

**end** $k$

**end** $m$

- updating the CD basis:

$$\mathbf{S}_n = (1-\tfrac{1}{i})\mathbf{S}_{n-1} + \tfrac{1}{i}\,\mathbf{S}_{n-1}\mathbf{H}_n$$

- updating the weight vector:

$$\mathbf{w}_n = \sum_{k=1}^{N_S} \frac{\mathbf{s}_{nk}^{H}\mathbf{v}_S}{r_{Ynkk}} \mathbf{s}_{nk}$$

**end** $n$

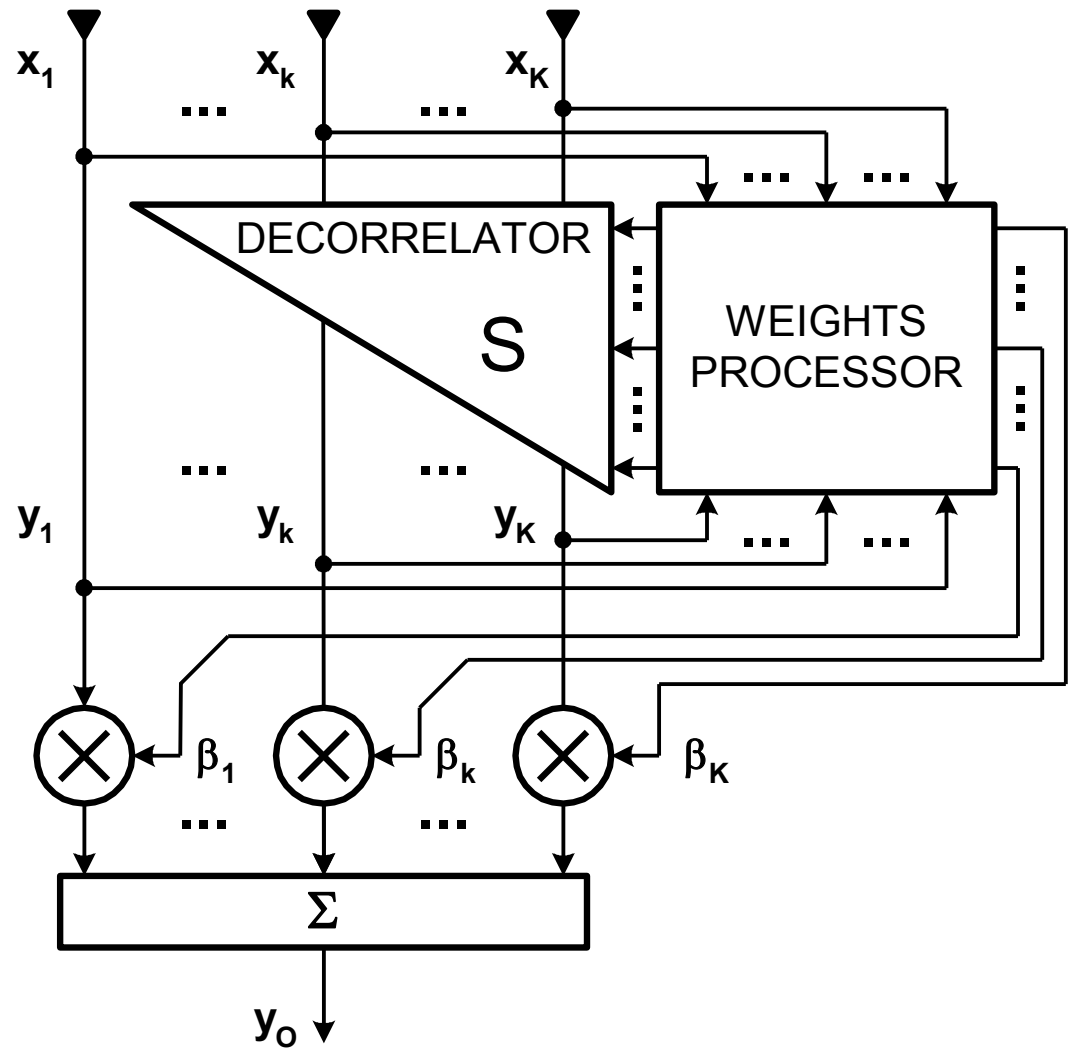

Fig. 1. Processing structure of space-time adaptive filtering based on feedback orthogonalization

A block diagram of the STAP filter implementing the new CDD based algorithm is presented in Fig.1.

From equations (11) to (14) and Fig.1 it becomes evident that the main feature of the newly designed STAP algorithm is in massive *feedback orthogonalization*. This technique effectively controls the depth and stability of adaptive decorrelation, and places stringent limits on the decorrelation weights thus suppressing accumulation and propagation of processing errors and faults to prevent the algorithm against explosive divergence.

## IV. SIMULATIONS

Simulations were performed in order to characterize the convergence performance of the new MSINR adaptive beamforming algorithm in a realistic situation for standard formats of fixed-point binary processing arithmetic.

In the simulations, we examine a uniform half-wavelength spaced linear antenna array with $K$ = 16 dipole elements. We assume the noise to be spatially isotropic zero-mean white Gaussian with unit variance.

The desired signal direction of arrival (DOA) is $\phi_S = 0^0$, its power is set to 10dB over the noise floor. All $M$ = 4 interfering signals are modelled by narrowband zero-mean Gaussian random processes, their powers and DOA's are presented in Table II.

TABLE III

| m | 1 | 2 | 3 | 4 |
|---|---|---|---|---|
| $p_m$ (dB) | 40 | 40 | 30 | 50 |
| $\phi_m$ ($^0$) | -12 | 10 | 18 | 23 |

For all simulation scenarios, the number of iterations was set to $n_F$ = 100 while the number of integrations required to estimate the output statistics in (13) was set to $n_I$ = 2. In order to demonstrate the new algorithm's stability of convergence, no ensemble averages were supposed.

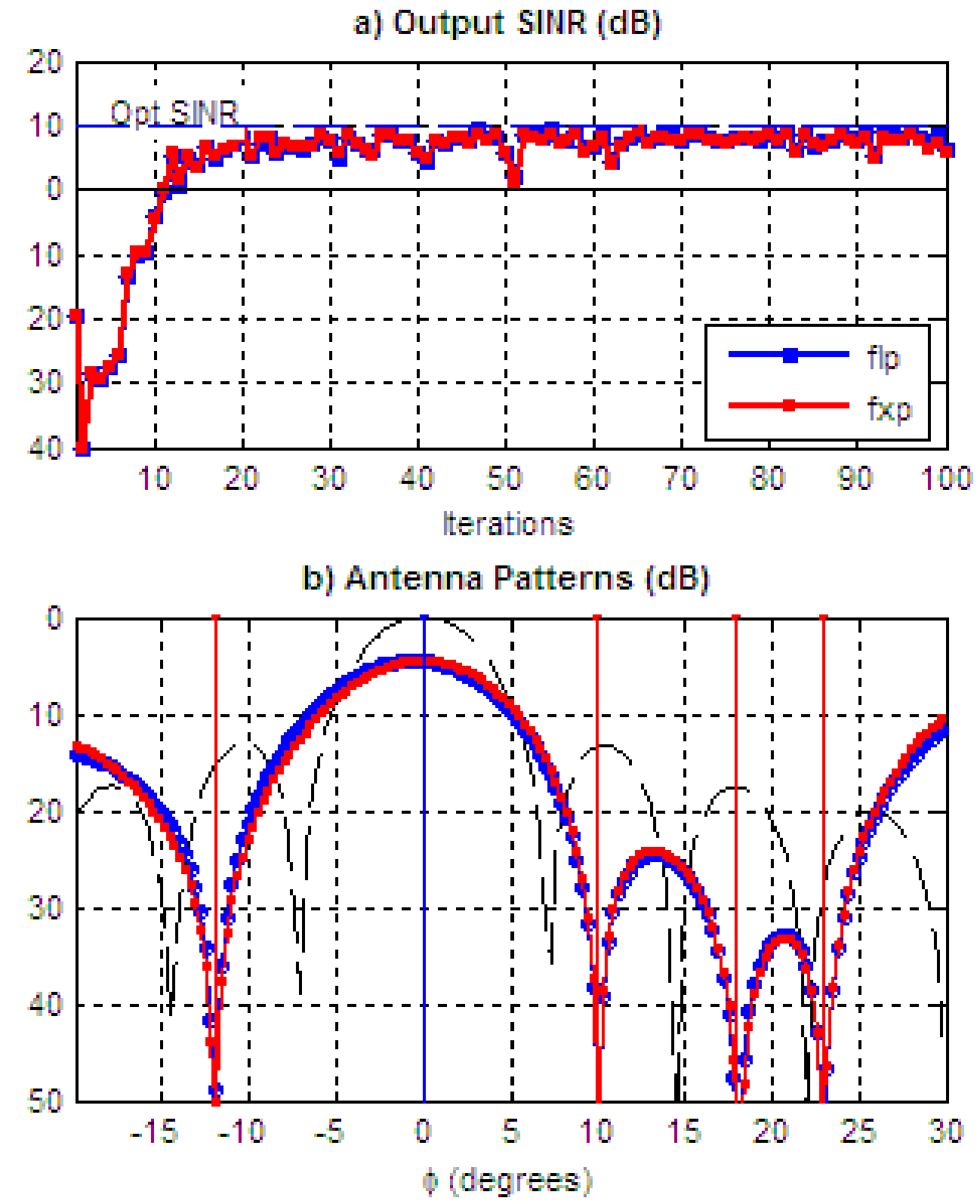

Fig. 2. Learning curves and antenna beam patterns for the 12-bit fixed-point implementation

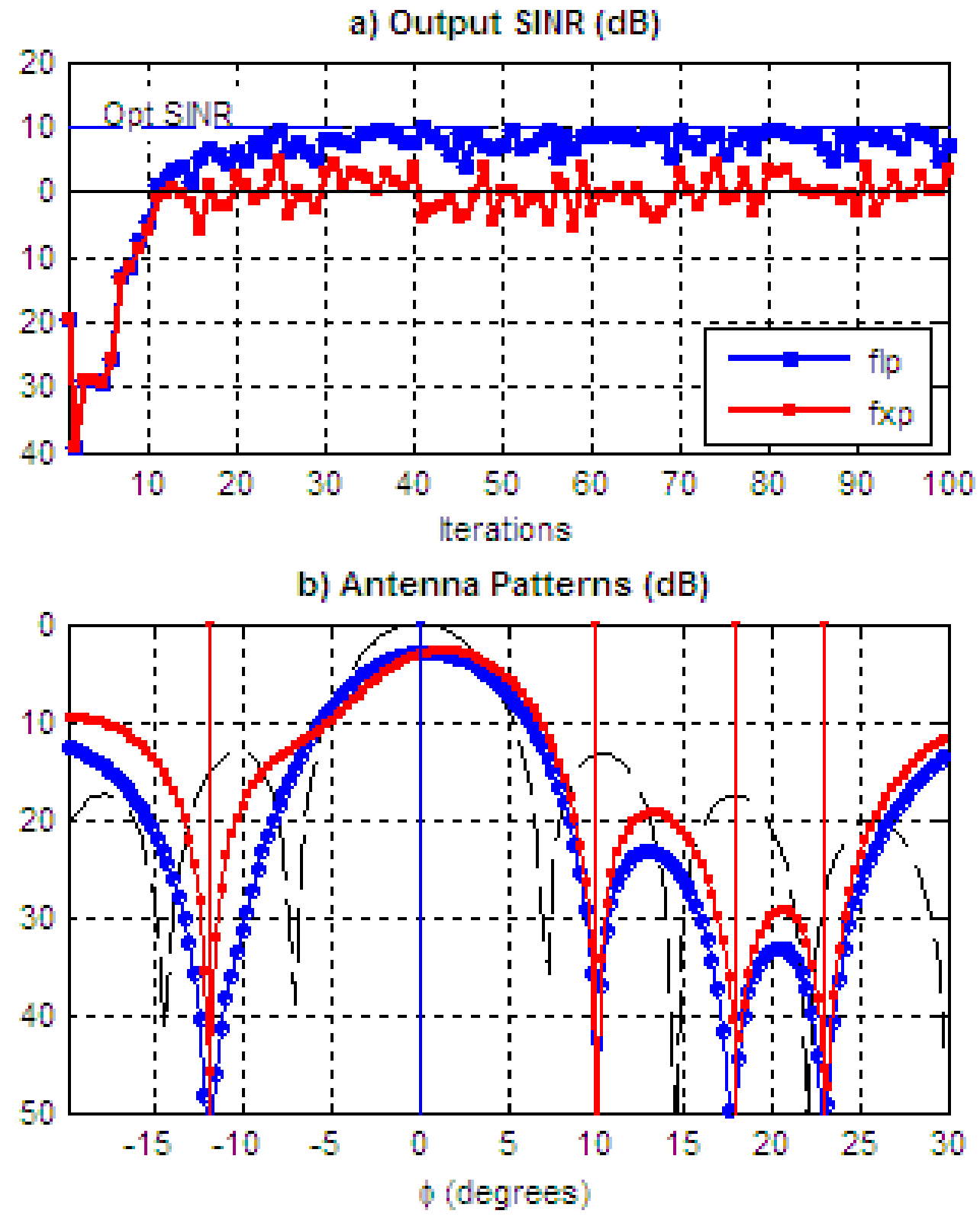

Fig. 3. Learning curves and antenna beam patterns for the 8-bit fixed-point implementation

The learning curves and antenna beam patterns obtained for the 12-bit and 8-bit fixed-point binary data/operation format are presented in Fig.2 and Fig.3, respectively. For comparison purposes, we also show the SINR learning curves and patterns obtained for 16-bit fraction floating-point format (blue color). The pre-adaptive beam pattern is marked with black color in the b) plots. The vertical red lines in the beam-pattern plots indicate the DOA's of the interferences.

## V. Conclusion

The CDD based approach applied to STAP algorithms designs has allowed us to synthesize a new fast-convergent adaptive filter with the highest immunity to numerical errors and faults that facilitates its real-time implementation on substantially faster and cheaper regular fixed-point processors.

Perfected numerical attributes of the new algorithm are provided by its inherent system of feedback orthogonalization which controls the performance of space-time decorrelation and suppresses propagation of processing errors and faults.

Due to capability of effective fixed-point implementation, this algorithm is expected to outperform most well-known STAP algorithms in terms of actual time of convergence.

To some extent, the newly designed CDD-algorithm can be considered as a seamless combination of a fast-convergent direct numerical method used to obtain a rough approximation to the optimum solution and a searching feedback algorithm targeted to improve the achieved approximation.